\documentclass[10pt,letterpaper,twocolumn]{article} %% two column, final layout

\usepackage{ol2}
\usepackage[draft,implicit=false]{hyperref}
\usepackage{amsmath}
\usepackage{gensymb,color}

\begin{document}

\twocolumn[

\title{Turbulence-induced persistence in laser beam wandering}

\author{Luciano Zunino,$^{1,2,*}$ Dami\'an Gulich,$^{1,2,3}$ Gustavo Funes,$^{4}$ and Dar\'io G. P\'erez$^{5}$}

\address{
$^1$Centro de Investigaciones \'Opticas (CONICET La Plata - CIC), C.C. 3, 1897 Gonnet, Argentina\\
$^2$Departamento de Ciencias B\'asicas, Facultad de Ingenier\'ia, Universidad Nacional de La Plata (UNLP), 1900 La Plata, Argentina\\
$^3$Departamento de F\'isica, Facultad de Ciencias Exactas, Universidad Nacional de La Plata (UNLP), 1900 La Plata, Argentina\\
$^4$Facultad de Ingenier\'ia y Ciencias Aplicadas, Universidad de Los Andes, Santiago, Chile\\ 
$^5$Instituto de F\'isica, Facultad de Ciencias, Pontificia Universidad Cat\'olica de Valpara\'iso (PUCV), Av. Brasil 2950, 23-40025 Valpara\'iso, Chile

$^*$Corresponding author: lucianoz@ciop.unlp.edu.ar}

\begin{abstract}
We have experimentally confirmed the presence of long-memory correlations in the wandering of a thin Gaussian laser beam over a screen after propagating through a turbulent medium. A laboratory-controlled experiment was conducted in which coordinate fluctuations of the laser beam were recorded at a sufficiently high sampling rate for a wide range of turbulent conditions. Horizontal and vertical displacements of the laser beam centroid were subsequently analyzed by implementing \textit{detrended fluctuation analysis}. This is a very well-known and widely used methodology to unveil memory effects from time series. Results obtained from this experimental analysis allow us to confirm that both coordinates behave as highly persistent signals for strong turbulent intensities. This finding is relevant for a better comprehension and modeling of the turbulence effects in free-space optical communication systems and other applications related to propagation of optical signals in the atmosphere.
\end{abstract}
\ocis{000.5490, 010.1290, 010.1300, 010.1330, 010.3310, 010.7060.}]

% 000.5490 --->>> Probability theory, stochastic processes, and statistics
% 010.1290 --->>> Atmospheric optics
% 010.1300 --->>> Atmospheric propagation
% 010.1330 --->>> Atmospheric turbulence
% 010.3310 --->>> Laser beam transmission
% 010.7060 --->>> Turbulence

\noindent
The centroid of a laser beam experiences multiple deflections when propagating through the Earth's turbulent atmosphere due to stochastic refractive-index fluctuations along the optical path. As a consequence of phase changes, due to turbulent eddies with dimensions larger than the beam diameter, the laser suffers displacements perpendicular to the original direction of propagation. This phenomenon is known as \textit{beam wandering} or \textit{spot dancing}. In a first approach, the wandering of the laser beam in the turbulent atmosphere could be considered as fully random~\cite{Churnside1990}. This means that the coordinate fluctuations of the laser centroid at a time are independent of those associated with previous instants. However, there exists some partial evidence that long-range correlations are present in the underlying temporal dynamics~\cite{Jorgenson1992,McGaughey1997,Perez2004,Perez2004a,Gulich2007,Funes2007,Funes2013}. It is clear that a better understanding of the fluctuating temporal behavior is essential for improving laser practical applications such as those related to tracking and communication purposes. Particularly, beam wander is considered the main cause of substantial signal losses at the receiver plane in free-space laser communication systems, degrading its data transmission quality and reliability, and limiting its performance~\cite{Kaushal2011}. Indeed, beam-wander mitigating control systems have been recently proposed for overcoming this drawback~\cite{Raj2014,Hulea2014}. Trying to shed some light on the laser beam wandering dynamics, in this Letter, we carefully analyze the temporal correlations in the recorded position of a laser beam after it propagates through an indoor laboratory atmospheric chamber. We conjecture that this laboratory-generated turbulence is representative of fully developed atmospheric turbulence. Consequently, it is able to emulate the main properties of the turbulent flows that affect the laser beam of a communication link. In fact, similar devices have been used for testing atmospheric turbulence effects in several contexts~\cite{Kaushal2011,Masciadri1997,Pors2011,Pereira2013,Zvanovec2013,Avramov-Zamurovic2014,Farias2015}. Horizontal and vertical components of the centroid position of the laser spot are measured as function of time with a position sensitive detector located at the end of the propagation path. With the aim to statistically characterize the temporal correlations of these data streams, a \textit{detrended fluctuation analysis} (DFA) is implemented. DFA is a robust technique for detecting dependence among samples in noisy non-stationary time series. The results obtained from this fractal analysis confirm that the displacements of the centroid of the laser are consistent with long-memory correlated stochastic dynamics for the stronger turbulent conditions, \textit{i.e.} when the turbulence effects can be resolved by the detector.

\begin{figure}[h]
\centerline{\includegraphics[width=8.5cm]{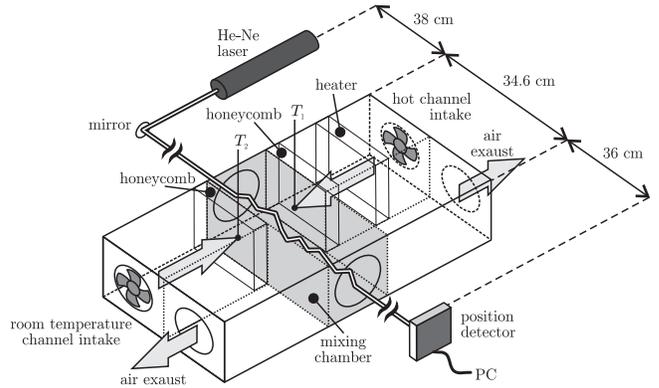}}
\caption{Schematic diagram of the laboratory experimental setup.}
\label{figure1}
\end{figure}

A conceptually simple experiment was performed in controlled conditions in which a laser beam propagates through artificial
turbulence---please see Fig.~\ref{figure1} for a schematic view of the optical setup. The wandering of the laser beam (10 mW HeNe Melles Griot Model 05-LHP-991) is detected by a position sensitive detector with an area of $1~\text{cm}^2$ (UDT SC-10 D). This detector measures the position of the centroid of the impinging laser beam with an accuracy of $2.5~\mu\text{m}$. Horizontal and vertical coordinates were recorded at 500 Hz---we have confirmed similar findings for higher sampling rates. For the purpose of having fully developed inertial turbulence at stable and statistically repeatable conditions, we employ a laboratory air turbulence generator, commonly called \textit{turbulator}, similar to the one originally proposed by Fuchs~\textit{et al.}~\cite{Fuchs1996}, and later enhanced by Keskin~\textit{et al.}~\cite{Keskin2006}. To simulate the atmospheric turbulence, two air fluxes at different temperatures are forced to collide in the chamber producing an isotropic mix between hot and cold air. The hot source is an electric heater controlled by changing the current passing through it. The thin laser beam propagates across almost $0.35$ m of turbulence in the mixing chamber. Air flow velocity is fixed because both fans operate at identical velocities so the turbulence characteristics are only due to the temperature difference. By increasing the temperature of the hot source different turbulent intensities can be produced. The indoor chamber offers the advantages of full system characterization and repeatability in a single turbulent layer. The strength of the artificial turbulence, quantified through the structure constant $C_n^2$, and the inner and outer scale were previously estimated following the procedure suggested by Masciadri and Vernin~\cite{Masciadri1997}. For such purpose, the variance of angle-of-arrival fluctuations of collimated laser beams as a function of the radius of different pupil masks passing through the turbulent layer is measured \cite[Fig.~1]{Masciadri1997}. By analyzing the averaging effect of the pupil sizes, and considering the effects of the inner ($l_0$) and outer ($L_0$) scales through the von K\'arm\'an spectrum, it is feasible to estimate the different turbulence parameters ($C_n^2$, $l_0$ and $L_0$) by the fit of the theoretical model~\cite[Eq.~(12)]{Masciadri1997} to the empirical variances. Finally, $C_n^2$ is expressed as a function of the temperature difference between hot and cold sources ($T_1$ and $T_2$, respectively, in Fig.~\ref{figure1}). Experiments with twelve temperature differences $\Delta T=T_1-T_2$ ranging from 5 to 180~\celsius~were carried out. Since turbulence is based on temperature gradient and air flow mixing, reference measurements were taken in two different conditions: fans on and fans off, both with the heater disconnected. In particular, measurements with the fans off can be considered as a background measurement that quantifies the electronic noise and room turbulence effects. It is worth noting here that the estimated structure constants are, at least, two or three orders of magnitude larger that those expected in outdoor experiments. Since the small turbulent path length in the mixing chamber, these higher turbulence strength are required to become detectable the laser centroid fluctuations.

DFA was introduced more than twenty years ago to characterize the fractal dynamics of a system from which a time series has been measured~\cite{Peng1994}. Being the most popular approach to detect the presence of long-term memory in data~\cite{Grech2013}, at present there are more than 1800 articles published on DFA and its applications according to the information extracted from the Scopus bibliographic database (accessed in March, 2015). However, specific applications in the optical field are scarce~\cite{Barille2006,Das2013,Funes2013,Zunino2014}. Briefly explained, the DFA method consist of five steps~\cite{Kantelhardt2001}. First, given a time series $S=\{x_t,t=1,\dots,N\}$, with $N$ being the number of equidistant observations, the cumulated data series $Y\left(i\right)=\sum_{t=1}^{i}\left(x_{t}-\langle x \rangle \right)$, with $i=1,\dots,N$ and $\langle x \rangle = \left(\sum_{t=1}^{N} x_t\right)/N$, is considered. In the second step, this profile is divided into $\left\lfloor N/s \right\rfloor$ nonoverlapping windows of equal length $s$ ($\left\lfloor a \right\rfloor$ denotes the largest integer less than or equal to $a$). A local polynomial fit $y_{\nu,\,m}\left(i\right)$ of degree $m$ is fitted to the profile for each window $\nu=1,\dots,\left\lfloor N/s \right\rfloor$ as the third step. The degree of the polynomial can be varied to eliminate constant ($m=0$), linear ($m=1$), quadratic ($m=2$) or higher order trends of the profile. Then, in the fourth step, the variance of the detrended time series is evaluated by averaging over all data points $i$ in each segment $\nu$, $F^{2}_m\left(\nu,s\right)=\left(1/s\right)\sum_{i=1}^{s}\left\{Y\left[\left(\nu-1\right)s+i\right]-y_{\nu,\,m}\left(i\right)\right\}^{2}$, for $\nu=1,\dots,\left\lfloor N/s \right\rfloor$. In the last step, the DFA fluctuation function is obtained by averaging over all segments and taking the square root, $F_m\left(s\right)=\left\{\left(1/\left\lfloor N/s\right\rfloor \right)\sum_{\nu=1}^{\left\lfloor N/s \right\rfloor}\left[F^{2}_m\left(\nu,s\right)\right]\right\}^{1/2}$. This procedure should be repeated for different values of the time scale $s$ in order to unveil the $s$-dependence of $F_m$. If the time series has long-range power-law correlations, $F_m\left(s\right)$ scales as
\begin{equation}
F_m\left(s\right)\sim s^H
\label{powerlaw}
\end{equation}
for a certain range of $s$. The Hurst exponent, \textit{i.e.} the scaling exponent $H$, is estimated by the slope of the best linear regression in a double logarithmic plot. It quantifies the long-range correlations embedded in the time series: when $H>1/2$, consecutive increments tend to have the same sign so that these processes are \textit{persistent}. For $H<1/2$, on the other hand, consecutive increments are more likely to have opposite signs, and it is said that the processes are \textit{anti-persistent}. $H=1/2$ is obtained for uncorrelated data~\cite[Chap. 9]{Feder1988}.

\begin{figure}[h]
\centerline{\includegraphics[width=9.5cm]{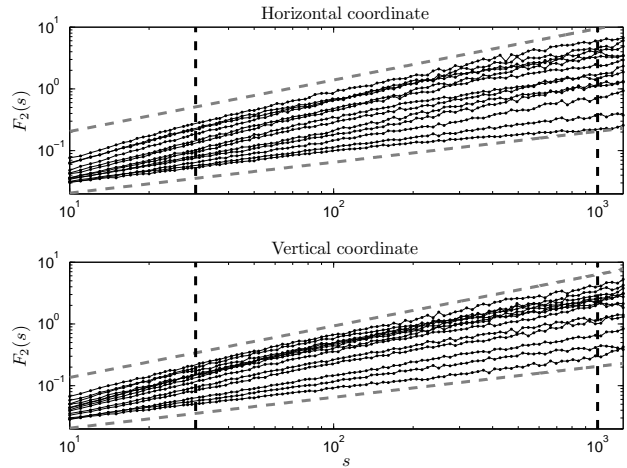}}
\caption{Fluctuation functions $F_2(s)$ as a function of the scale $s$ for the horizontal (upper plot) and vertical (lower plot) coordinate fluctuations for the different turbulent conditions. Results obtained for one particular realization of the twenty one recorded are depicted. A detrending polynomial of order $m=2$ and 96 different scales $s \in [10,N/4]$ equally spaced in the logarithmic scale were employed in the DFA implementation. $C_n^2$ increases from bottom to top. The slope of the
best linear fit obtained for each one of these fluctuation functions is the Hurst exponent estimator. Vertical dashed lines indicate the range in which the linear fit for estimating the Hurst exponent is performed. Straight (gray dashed) lines with slopes 1/2 (bottom line) and 5/6 (top line) are also shown as references. The behavior observed is representative for the whole data set and similar results are obtained for other detrendings ($m=1$ and $m=3$). The significant similarity between horizontal and vertical fluctuation functions can be attributed to isotropy of the artificially generated turbulence.}
\label{figure2}
\end{figure}

\begin{figure}[h]
\centerline{\includegraphics[width=9.5cm]{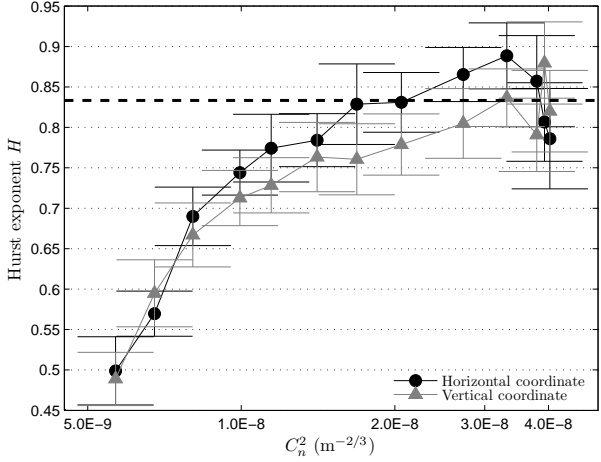}}
\caption{Hurst exponent estimated values for the horizontal and vertical coordinates of the laser beam centroid as a function of the turbulence strength. Mean and standard deviation of the twenty one realizations for each $C_n^2$ value are plotted. The theoretical expected value for the Hurst exponent within the Kolmogorov model ($H=5/6$) is indicated (horizontal black dashed line). A second order DFA algorithm with linear fitting range $[30,1000]$ was implemented. Behaviors obtained for other orders ($m=1$ and $m=3$) are very similar.}
\label{figure3}
\end{figure}

\begin{figure}[h]
\centerline{\includegraphics[width=9.5cm]{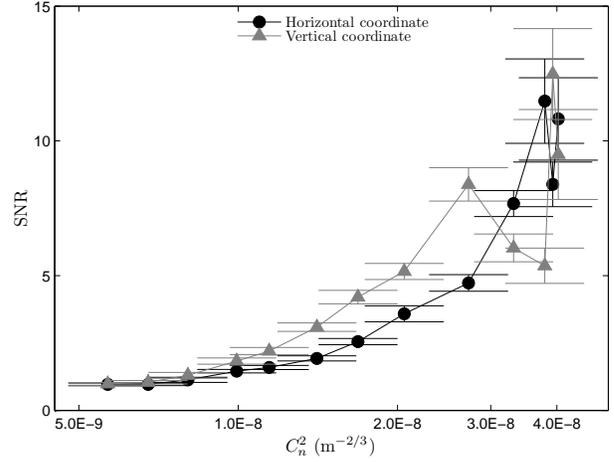}}
\caption{Signal-to-noise ratio for the horizontal and vertical coordinates of the laser beam centroid as a function of the turbulence strength. Mean and standard deviation of the twenty one realizations for each $C_n^2$ value are plotted.}
\label{figure4}
\end{figure}

Twenty one independent realizations of 5000 coordinate points were obtained for each turbulent condition. Afterwards, the DFA analysis was performed for both horizontal and vertical coordinates. Figure~\ref{figure2} shows the fluctuation functions obtained by implementing a DFA analysis with a detrending polynomial of second order ($m=2$) for the different turbulent strengths. Representative results for one particular realization for each turbulent condition are depicted, with $C_n^2$ increasing from bottom to top. A well-defined power-law behavior is concluded with higher slopes for the stronger turbulent intensities. Mean and standard deviation (over the twenty one independent realizations) of the Hurst exponents estimated in the range $s \in [30,1000]$ (vertical dashed lines in Fig.~\ref{figure2}) as a function of $C_n^2$ for both coordinates are plotted in Fig.~\ref{figure3}. In order to better interpret the results obtained for the Hurst exponent, Fig.~\ref{figure4} shows the signal-to-noise ratio (SNR) for the different turbulent conditions. SNR was estimated as the variance of the signals at turbulent states relative to the value associated with the background measurements. From Fig.~\ref{figure4} it is concluded that higher $C_n^2$ are needed to fully resolve the turbulence effects in the turbulator with the implemented detection system. 

For the reference measurements the detector is unable to resolve the position differences, and a fully uncorrelated electronic noise associated with the detector is measured. Consequently, the Hurst exponent is near 0.5 as expected. As the turbulent strength increases, the SNR is larger than one (please see Fig.~\ref{figure4}) and the detector begins to discern the turbulence influence. Simultaneously, the Hurst exponent shows an increasing behavior, for both horizontal and vertical coordinates, saturating at a value close to 5/6 for the higher $C_n^2$ values (please see Fig.~\ref{figure3}). Results obtained allow us to confirm that turbulence introduces memory effects in time series wandering because highly persistent dynamics are clearly concluded for the stronger turbulence intensities. Moreover, the similarity between horizontal and vertical estimated Hurst exponents confirms the isotropy of the turbulence within the laboratory chamber. It is worth mentioning here that $H=5/6$ has been originally proposed theoretically for the turbulence-degraded wavefront phase within a Kolmogorov model~\cite{Schwartz1994}, and very recently confirmed experimentally for the angle-of-arrival fluctuations of stellar wavefronts propagating through atmospheric turbulence~\cite{Zunino2014}.

Summarizing, a persistent stochastic fractal behavior is clearly concluded from the DFA analysis of laser beam wandering in laboratory-generated turbulence. Estimated Hurst exponents, for both coordinates, converge to a value close to 5/6---the theoretical value associated with Kolmogorov turbulence---for the stronger turbulent conditions. Outdoor similar wandering measurements for horizontal paths in different moments of the day are being planned for the near future in order to confirm the presence of memory effects in real atmospheric channels.

This work was partially supported by Consejo Nacional de Investigaciones Cient\'ificas y T\'ecnicas (CONICET), Argentina; Universidad Nacional de La Plata (UNLP), Argentina; Comisi\'on Nacional de Investigaci\'on Cient\'ifica y Tecnol\'ogica (CONICyT, Chile) (FONDECYT 1140917) and Pontificia Universidad Cat\'olica de Valpara\'iso (PUCV, Chile) (123.731/2014).

\pagebreak

\end{document}